# ATOMISTIC SIMULATION OF COMPRESSION WAVE PROPAGATION IN NANOPOROUS MATERIALS

Mark Duchaineau, Alex V. Hamza, Tomas Diaz De La Rubia, Farid F. Abraham

Lawrence Livermore National Laboratory
Livermore, CA 94550

## Abstract

We have developed a method to simulate behavior of nanoporous materials in a molecular dynamics code. The nanoporous solid is "produced" via a spinodal decomposition of a material brought from a supercritical fluid into the two phase (liquid-vapor) region and then quenching and freezing the liquid into an interconnected nanoporous solid. We have simulated, at the atomic level, compression in crystal/nanoporous configurations, demonstrating that this is a powerful technique for studying the equation-of-state of cold and warm dense matter. By performing compression simulations relevant to high energy density physics experiments, we have been able to elucidate experimental measurement by identifying governing microscopic mechanisms.

## Introduction

Predicting compression wave propagation in porous materials is critical in understanding seismic waves, advanced recovery of fuels in porous media; in high energy materials processing to form novel materials; and in porous solid propellants. An accurate description of porous solids is central to understanding filtration, drug delivery, and flow in porous media. The development of robust simulation capabilities to predict the behavior of matter under extreme conditions is of central importance to high energy density science community. There is a premium in predicting the thermodynamic and mechanical response of materials subjected to shock-induced and shockless compression, either with mechanical impactors (gas gun) or laser driven ablators. The emergence of high-energy laser systems, such as the National Ignition Facility (NIF), affords the opportunity to create and interrogate states of matter under unprecedented extreme conditions of pressure, temperature, strain, and strain rates. However, what is lacking is a fundamental understanding of the relationship between the microphysics of the materials properties of the drive impactor ("reservoir") and the target materials. An atomic level simulation effort can serve as the basis to provide a systematic scientific underpinning to relate the microstructure of the reservoir impactor to the pressure profile on the target. The results of these ultra-scale atomic level simulations can be significant in improving the predictive capabilities of continuum-level simulations of laser-driven shock compression experiments and in engaging the scientific community in this emerging field. It is critical to be able to calculate the leading edge of the ramp wave, as this sets the initial adiabat from which the high pressure loading occurs. State-of-the-art hydrodynamic codes cannot calculate the leading edge correctly. We will show that our large scale molecular dynamics (MD) simulations succeed in capturing the leading edge of the ramp wave, strengthening the scientific underpinnings for high pressure compression experiments proposed for NIF and other high energy density platforms.

Planetary interiors are high pressure, relatively low temperature environments. In order to study matter under these conditions a laser-based platform is being developed. In the traditional laser-based experiment a strong shock is launched in a material which instantaneously increases both the temperature and pressure. In most cases the sample melts. Recent work by Remington and co-workers [1, 2, 3] has demonstrated the possibility of a quasi isentropic compression on a laser-based platform. The intense laser pulse drives a reservoir which unloads across a gap onto the sample of interest. Multi-Mbar (100's GPa) pressures can be achieved. Similar experimental platforms have been produced with other drivers, such as gas guns, magnetic flyers on Z-pinch, and high explosives. Higher power lasers such as the NIF laser can drive samples to over 10 Mbar pressure. However the rise time of the pressure pulse can shorten to a shock. Very recently Smith *et al.* [1] have shown it may be feasible to lengthen or stretch the rise time to these high pressures by using a graded density reservoir. In this paper we use molecular dynamics simulations to investigate the pressure loading of a solid sample to elucidate the effects of the compression wave propagating through the graded density reservoir. These simulation techniques provide a tool to design and tailor the pressure pulse in an isentropic compression of a solid.

In addition a well known laser-based technique for measuring the equation-of-state of solids at high pressure (Mbar) involves comparing the measured transit of a shock wave through a reference material with a well-established equation-of-state and the solid of interest [4]. This technique has been extended to porous materials as well [5]. However, the nature of the shock or compression wave in the porous material is not well known. These simulations reported here also reveal the behavior of the compression wave traveling through porous materials which are integral to the equation-of-state measurements.

Our systems of interest are composite crystal/nanofoam sandwiches. Low density porous materials are used in many high energy density physics experiments, because they provide a convenient method to independently vary density without changing the atomic number. Of particular concern is the effect of pore size on the hydrodynamic behavior during shock loading and unloading into vacuum, or on to other porous solids and non-porous solid materials. It is known that large pores (>10 - 50 microns) can imprint structures on other materials and produce hydrodynamic instabilities. What is not known is at what pore size the hydrodynamic instabilities are mitigated and imprinting no longer occurs. It is believed that pore sizes between 10 and 100 nanometers do not seed hydrodynamic instabilities and do not imprint on other materials. Simulation is testing this.

We have constructed physical models and developed the simulation and analysis programs with supporting visualization tools for application to the study of shockless compression by large-scale computation. Our beginning effort was the realization that a metallic nano-foam could be created on the computer by rapidly quenching a high-temperature fluid undergoing spinodal decomposition during phase separation. This has led to a study of the mechanical properties of metallic copper foams of various porosities and our current interest in studying their dynamic failure by shock compaction. We first discuss the formation of a porous solid on the computer.

## Simulating Nanoporous Solids

Recently Zhao et al. [6] have simulated a porous solid by introducing four 4 nm diameter spherical voids in an otherwise fully dense solid. The isolated, close cell porosity simulated provides only part of the picture, since the voids do not directly interact with each other. We conceived of a computational procedure for "creating" a nanoporous solid which is tailored for simulation. The technique uses the dynamic phase transition process, spinodal decomposition, and is depicted in Fig. 1. In this figure, we see the familiar nucleation process being initiated by the appearance and growth of micro-droplets when the supercritical fluid is quenched into the two-phase coexistence region of vapor and liquid. However a quench into the central region of coexistence gives rise to phase separation by spinodal decomposition represented by a "porous fluid network" of high and low density regions which will rapidly break into droplets because of surface tension [7]. Before this breakup occurs, we quench this highly interconnected phase separating fluid to form a frozen porous solid.

The initial porous sample is small and amorphous. It has irregular filament shapes due to the physical simulation process by which the sample was produced. For studies of the material response of pore structures under various conditions, it is desirable to construct pore structures at larger length scales, with controlled atomic arrangements (e.g. perfect crystal, grains, etc), and with specified density profiles or filament cross-sections. We next describe a topological filtering approach to analyze and synthesize such designed pore structures using the original small, amorphous sample as a template.

Overall the analysis and synthetic processing involves these steps:

1. Input an initial set of atom positions from the spinodal decomposition process;
2. Make a proximity field of original atoms on a regular grid;
3. Compute a signed distance field relative to the solid/void interface surface;
4. Perform a topology-preserving surface propagation from the interface surface to produce a topologically "clean" distance field and curved skeleton of the pore filament structure;

5. Compute a distance field from the curved skeleton, re-scaled to produce uniform density profiles;
6. Use the re-scaled distance field to "carve out" atoms with identical pore filament topology as the original sample, but with specified scales, density profiles, and atomic arrangements.

These six steps are illustrated in Figure 2.

A proximity field is a scalar-valued function sampled on a cubic grid that fills the periodic domain of the spinodal decomposition process that produced the original pore structure. The grid spacing should be fine enough to resolve small filaments and voids. Spacing equal to the average inter-atomic distance in a bulk crystal is sufficient. The proximity field is the sum of Gaussian kernel functions centered on each atom. For each atom in the original sample, all grid vertices within a specified multiple of the inter-atomic distance of the atom center are evaluated (a factor of 4.0 is safe). The distance between these vertices and atom center is computed, and the field value for the vertex is updated by adding the Gaussian function of this distance. The Gaussian is truncated carefully to have zero value and first derivative at the cutoff distance. The original atom positions are illustrated in Figure 2a, with the proximity field in Figure 2b. An appropriate proximity contour forms an adequate representation of the solid/void interface.

Next, a signed distance field is computed for the solid/void interface of the original sample, see Figure 2c. The distance field is the closest distance to the boundary surface between the void and the solid ligament and that distance is positive or negative depending on whether the point is in the ligament or in the void. This is also a scalar field sampled on the same cubic grid as the proximity field. Every vertex of the volume grid stores the shortest distance to the volume/solid interface surface. Distances in the void regions are negated, while distances inside the solid remain positive. This distance field is computed using a queue-based front-propagation method described in Laney et al [8].

The main analysis step is the computation of the centerline "skeleton" of the solid pore filaments. A front propagation method computes this field in a series of "onion layers" growing inward and outward from the reference interface surface. The rate of propagation perpendicular to the front is typically constant (gradient magnitude 1.0), except when a topological change is imminent locally, which causes the propagation to slow down (perhaps temporarily). Details of this construction process are given in Gyulassy et al [9]. At the end of this construction process, a set of line segments representing the centerlines of the solid filaments is output, as shown in Figure 2d. Due to the irregularities in the solid/void reference surface, these centerlines can be noisy. A variational smoothing process is used on the lines, with a force term in the distance field gradient direction to keep the lines centered. In analogy to a physical process the skeleton is determined by isotropic uniform removal of layers of the filament until it reaches a one dimensional representation. One caveat to this process is filaments cannot break or new pores cannot be formed. The resulting smoothed centerlines are shown in Figure 2e.

Two additional fields are constructed from the curved skeleton (i.e. from the centerlines of the solid filament network). The first is a conventional distance field from the curved skeleton. This is useful for distance-based analysis and in synthesis of pore structures with specified radii of filament cross sections. The second field is similar, but where each (X, Y) cross section is normalized so that thresholding may be performed to get specified relative densities from 0% to 100%. Figure 2f shows an example of a synthesized porous solid with constant density for each Y-axis cross section. Note that constraining a solid to constant density profiles typically requires variable ligament radii.

A movie of the nanofoam structure is included in the supplementary informations (see movie S1). For the three-dimensional pore structure used in this study, a solid/void interface was determined (shown as a green surface in the movie), along with the solid filament centerlines (shown in red). The centerlines are used in analysis to determine structural characteristics of the porous materials during shocks and/or deformations, to determine line densities, length distributions, filament displacements, breaks, and merges. The centerlines and related distance fields enable the synthesis of porous solids for simulation.

## Constructing the Target Sandwich

Our current effort is to probe the dynamical response of the nanoporous copper metal foam by simulating its behavior when impacted by a shock. Figure 3 shows three different sandwiched configurations of copper crystals and foam which are of interest for equation-of-state measurements, lattice dynamics, materials strength, and isentropic compression experiments (see references [3, 10]. Other materials of current interest are gold, silica, and carbon. We will refer to the configurations as sandwiches. In all cases, a sandwich is pushed by a piston from the left at a speed that generates a shock with desired characteristics in the left-hand solid. The shock impacts the foam, creating a "compaction wave" which eventually impacts the right-hand neighboring crystal after exploding across a vacuum gap (middle and last picture). In this paper, we will now discuss simulations for the second sandwich structure in Figure 3.

The structure of our two simulated sandwiches, A and B, are drawn to scale and presented in Figure 4. For sandwich A, the nanoform section has a constant density equal to 25% bulk throughout its length and filament size of approximately 4 nanometers. For sandwich B, the nanofoam section has a linear variable density from 40% bulk to 10% bulk as depicted in the figure. The piston speed is 3 kilometers/second.

Our simulation tool is computational molecular dynamics [11, 12]. Molecular dynamics predicts the motion of a given number of atoms governed by their mutual interatomic interaction, and it requires the numerical integration of the Newton's equations of motion, *"force equals mass times acceleration or F = ma."* A simulation study is defined by a model created to incorporate the important features of the physical system of interest. These features may be external forces, initial conditions, boundary conditions, and the choice of the interatomic force law. We have discussed the chosen model system. Our choice of the force law is the Voter-Chen [13] embedded atom potential describing copper.

## Shock Loading the Nanofoam Sandwich: Results & Analysis

Figure 5 shows the pressure profiles for the sandwich for the time interval from 30 to 50 ps at every 5 ps. From this type of simulation data the pressure as a function of time in crystal copper after the vacuum gap can be determined. Figure 6 shows this pressure versus time history for the crystal copper loaded by a porous copper reservoir compressed and expanded across a vacuum gap. A comparison of the constant porosity reservoir with the variable porosity reservoir is made. Focusing on the peak pressure due to the gas loading of the full density sample across the vacuum gap (~12 nm into the full density copper solid), we observe that the pressure rise is approximately the same in both the constant porosity and variable porosity cases. There are, however, subtle differences which are also observed in the experiment [1]. The length and time scales are different in the simulation and experiment; however, the velocity scaling is fairly good for the compaction wave in the porous material (about a factor of 2). In the experiment there is a 60 micron thick variable density sample with a 300 micron gap which gives a 17 ns rise time in the pressure. In the simulation ~120 nm thick variable porosity material with 60 nm vacuum gap gives a 17 ps rise. Both the simulation and the experiment show that the pressure rise is more gradual for the variable porosity case (experiment: rise time increases by 4ns, simulation: rise time increases by 2.5 ps). The experimental comparison between constant versus variable porosity reservoirs does not control for the mass of the reservoir eventually pushing on the sample; thus for the same laser drive the pressure reached in the variable porosity case is lower because there is less mass. The simulation comparison does control for the mass of the reservoir, i.e. the mass of the reservoirs in the variable and constant density cases is the same.

An important insight gained from the simulation not previously considered is the hot gases that race ahead of the compression wave in the porous media and across the vacuum gap. The racing ahead of hot gasses is similar to the formation of jets in hydrodynamics experiments. In both the constant and variable porosity cases the hot gases are the first material to load the sample of interest. The hot gases race ahead with some memory of the porous structure, since the pores allow the gases to move forward unimpeded, but the ligaments stop or slow the forward motion the gases. The memory of the structure may lead to possible imprinting on the sample. Another important insight gained from the simulation is the compaction of the porous solid as the compression wave passes must be considered when predicting the performance of the reservoir in the experiment. The "accordion" effect in the compaction of the porous solid produces and intermediate speed for the compression wave between the piston speed and the shock speed in a solid of the same density as the porous solid.

Recent experimental results aimed at measuring the equation-of-state of 14-20% relative density nanoporous (sub 500nm pore size) copper have measured the propagation of a compression wave through the nanoporous copper at ~10-15Mbar pressure [14]. The experiment is performed at the OMEGA laser in Rochester, NY. The laser energy impinges on an ablator, in this case brominated polystyrene. A shock wave is launched into the polystyrene which is attached to a thin (40 micron thick) aluminum reference material. The equation of state is well known for the reference aluminum. The copper foam is bonded to the aluminum with a thin (~1 micron) silver solder layer. Continuity at the interface gives the piston or particle velocity in the copper foam. The wave propagation velocity through 15% relative density nanoporous copper was 34 km/s at ~10 Mbar pressure and 40 km/s at ~15 Mbar. The current simulation results give a comparable compression wave velocity (see figure 7). The interference measurements used for the shock velocity determination are not sensitive to the "hot gases" racing ahead in the simulation, since the only the uniform compression wave break out is observed. Movies of the compaction wave passing through the nanoporous copper are available in the supplementary information (see movies S2, S3, and S4).

# Conclusion

The atomistic simulations employed here provide insight into the microscopic mechanisms of shock propagation through porous solids and the loading of the samples across a vacuum gap from the porous solid. The hot gasses racing ahead of the compression wave shown in the simulation will lead to better design of experiments to measure high pressure properties of solids on high energy density platforms. Another important discrepancy between continuum models and experimental observation is that the shock propagation speed in porous solids is slower than predicted (see figure 7) [15]. The accordion effect of pressing filaments and hot gas together leads to a speed intermediate between the piston speed and the continuum model-predicted shock speed in a crystal solid (of the same density as the porous material).

**References**

[1] R. F. Smith, *et al.*, "Graded-density reservoirs for accessing high stress low temperature material states," Astrophysics and Space Science **307** (1-3): 269-272 (2007)

[2] K. T. Lorenz, *et al*., "Accessing ultrahigh-pressure, quasi-isentropic states of matter," Physics of Plasmas **12** (5): Art. No. 056309 (2005).

[3] J. Edwards, *et al*., "Laser-driven plasma loader for shockless compression and acceleration of samples in the solid state," Physical Review Letters **92** (7): Art. No. 075002 (2004).

[4] P. M. Celliers , G. W. Collins, D. G. Hicks, J. H. Eggert, "Systematic uncertainties in shock-wave impedance-match analysis and the high-pressure equation of state of Al," Journal of Applied Physics **98** (11): Art. No. 113529 (2005).

[5] J. E. Miller, T. R. Boehly, J. H. Eggert, "Equation-of-State Measurements in $Ta_2O_5$ Aerogel," Proceedings of International Conference on Shock Compression of condensed matter, to be published.

[6] S. Zhao, T. C. Germann, and A. Strachan, "Molecular dynamics simulation of dynamical response of perfect and porous Ni/Al nanolaminates under shock loading," Physical Review B **76**, 014103 (2007)

[7] F. F. Abraham, M. R. Mruzik, G. M. Pound, "EXPERIMENTAL-VERIFICATION OF THE SPINODAL CURVE," Journal of Chemical Physics **70** (5): 2577-2579 (1979).

[8] D. Laney, M. Bertram, M. Duchaineau, N. Max, "Multiresolution Distance Volumes for Progressive Surface Compression", Proceedings of the First International

Symposium on 3D Data Processing, Visualization, and Transmission, Padova Italy, June 19-21, 2002, pp. 470-479.
[9] A. G. Gyulassy, *et al.*, "Topologically Clean Distance Fields", IEEE TVCG (Visualization 2007), in press.
[10] R. W. Lee, D. Kalantar, and J. Molitoris, "Warm Dense Matter: An Overview," UCRL-TR-203844, April 29 2004.
[11] F. F. Abraham, "Computational Statistical-Mechanics Methodology, Applications and Supercomputing," Advances in Physics 35, 1 (1986).
[12] M. P. Allen and D. J. Tildesley, Computer Simulation of Liquids (Clarendon Press, Oxford, 1987).
[13] A. F. Voter, "Embedded atom method potentials for seven FCC metals :Ni, Pd, Pt, Cu, Ag, Au, and Al," Technical Report LA-UR 93-3901, Los Alamos National Laboratory, 1993.
[14] R. H. Page, Private Communication.
[15] R. Macri and T. Dittrich, Private Communication.

Figure Captions:

Figure 1: Making a nanoporous solid by spinodal decomposition

Figure 2: The steps in analysis and synthesis are shown starting in the upper left: (a) the original small amorphous sample of atoms; (b) the proximity field generated by summing a Gaussian blur kernel per atom; (c) a signed distance field with dark inside and light outside the in/out boundary contour of the proximity field; (d) the topologically equivalent curved skeleton (i.e. filament centerlines) for the distance field; (e) the centerlines after variational smoothing with a force term pushing along the distance field gradient; (f) a synthesized porous solid with constant cross-section densities for all Y-axis slices.

Figure 3. Sandwich configurations considered in this simulation study are shown. A solid crystal of copper begins each sandwich on the left hand side. Porous copper is next to the crystal copper on the right. On the top panel crystal copper completes the sandwich. In the middle panel a vacuum gap separates the porous copper from the copper crystal on the right. In the bottom panel a vacuum gap separates porous copper from porous copper attached to a copper crystal.

Figure 4. A schematic of the middle configuration from fig. 3 for two cases, sandwich A and sandwich B is depicted. Along with the schematic of the two sandwiches is also the density profile. In sandwich A the density of the porous copper is constant. In sandwich B the density of the porous copper decreases from 40% to 10% relative density from left to right. The length of the sandwich is 3029 Å and the width is 328 Å.

Figure 5. Plots of the pressure profile in the sandwich are shown for times from 30 to 50 ps in 5 ps intervals. From these types of profiles the data in figure 6 is derived.

Figure 6. The pressure as a function of time at a position just inside the copper crystal after the vacuum gap is shown. The pressure rises in the copper crystal due to loading by the porous copper coming across the vacuum gap. The pressure at the peak due to the loading occurs at ~100 GPa and at 40-45 ps.

Figure 7. The logarithm of compaction front speed in the porous copper versus the logarithm of piston speed is shown. A linear fit on the log-log scale to the simulation results is also displayed. The results from the laser driven experiment are shown as well [12]. For comparison the predictions to the hydrodynamic continuum calculation are also shown [13].

Movie S1. Movie of the nanoporous foam structure formed by the procedure described in the text.

Movie S2. Movie of the simulated response of the nanoporous copper (15% relative density) driven by piston speed of 2 km/s.

Movie S3. Movie of the simulated response of the nanoporous copper (15% relative density) driven by piston speed of 4 km/s.

Movie S4. Movie of the simulated response of the nanoporous copper (15% relative density) driven by piston speed of 5 km/s.

Figure1.

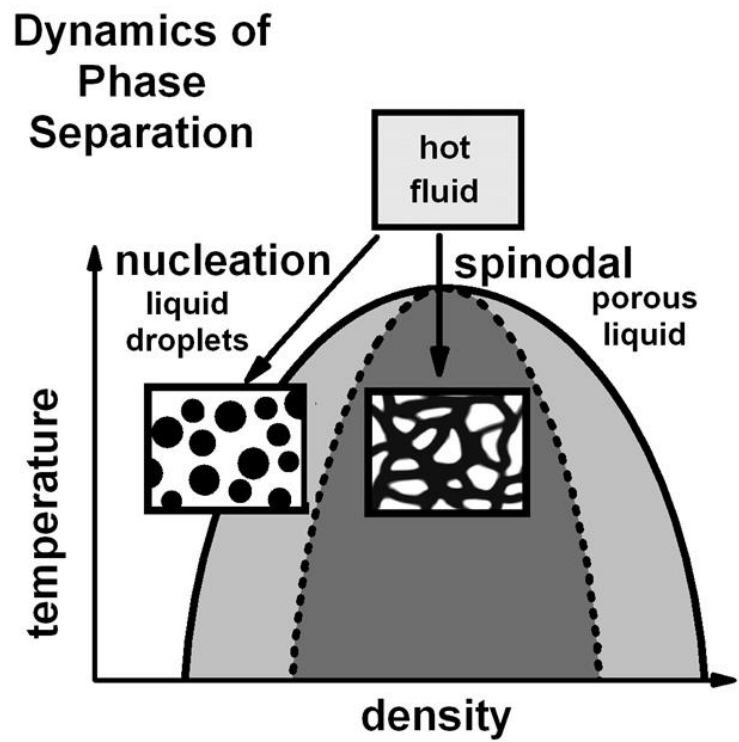

Figure 2

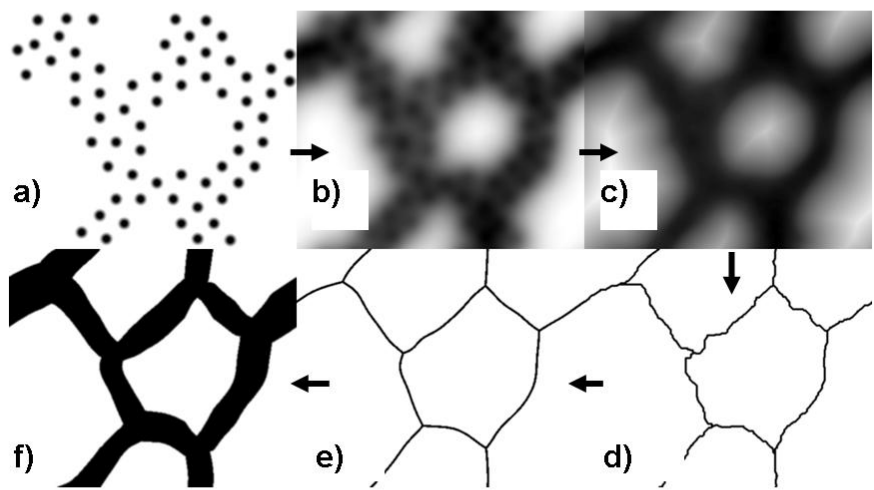

Figure 3.

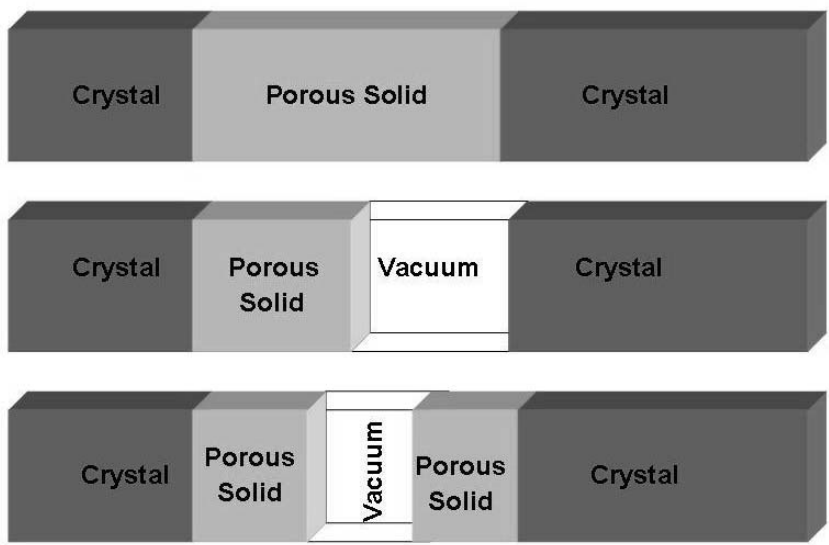

Figure 4.

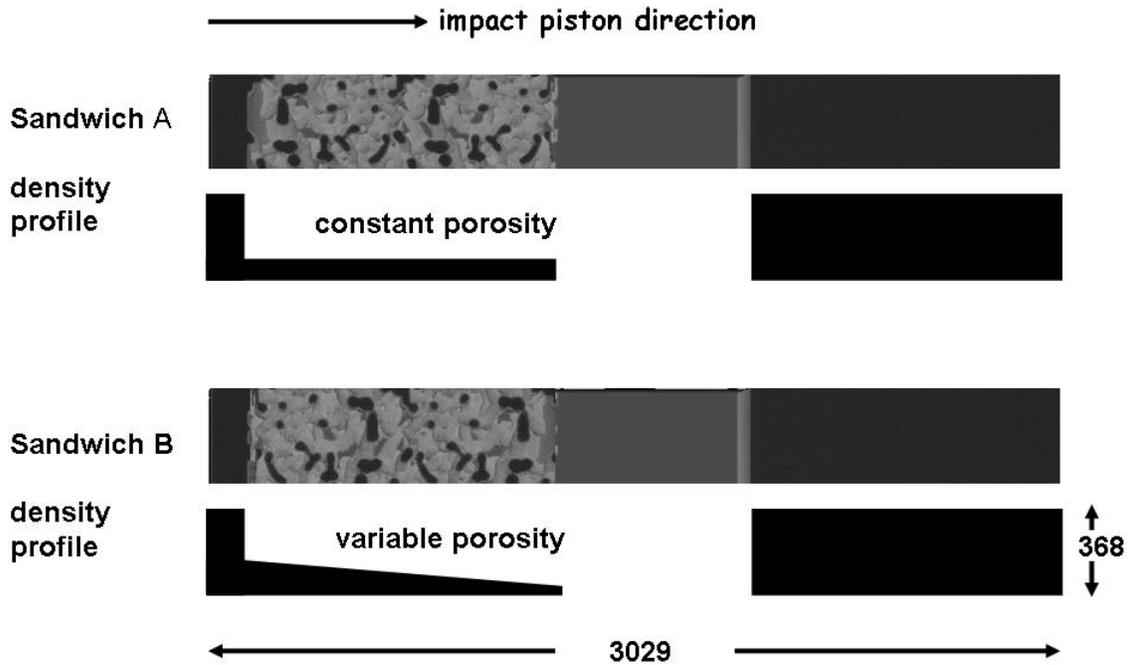

Figure 5.

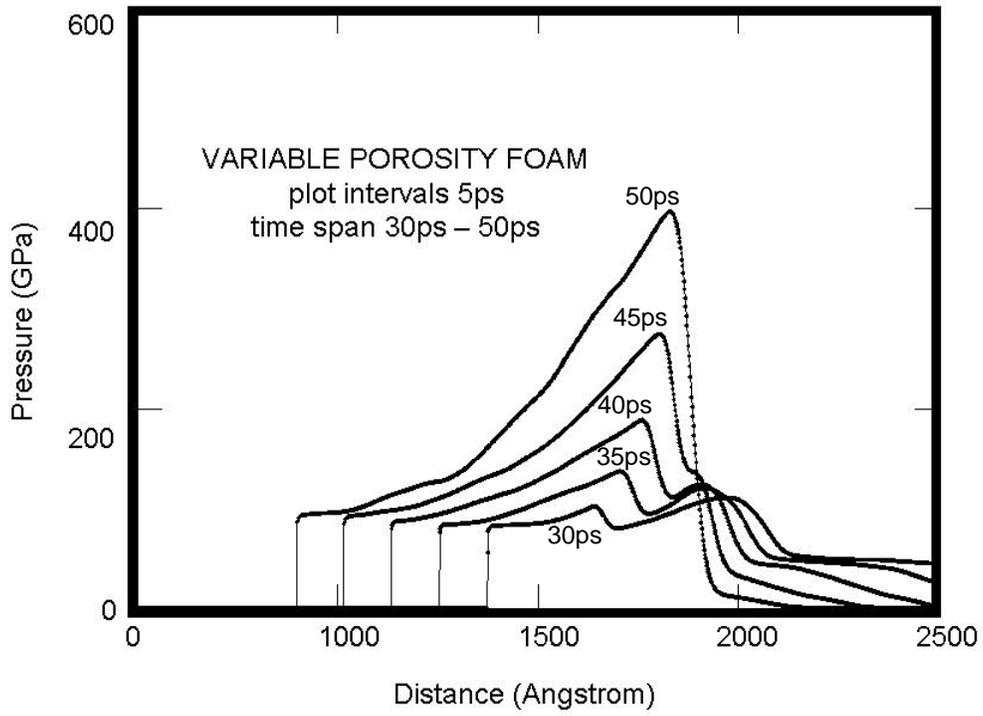

Figure 6.

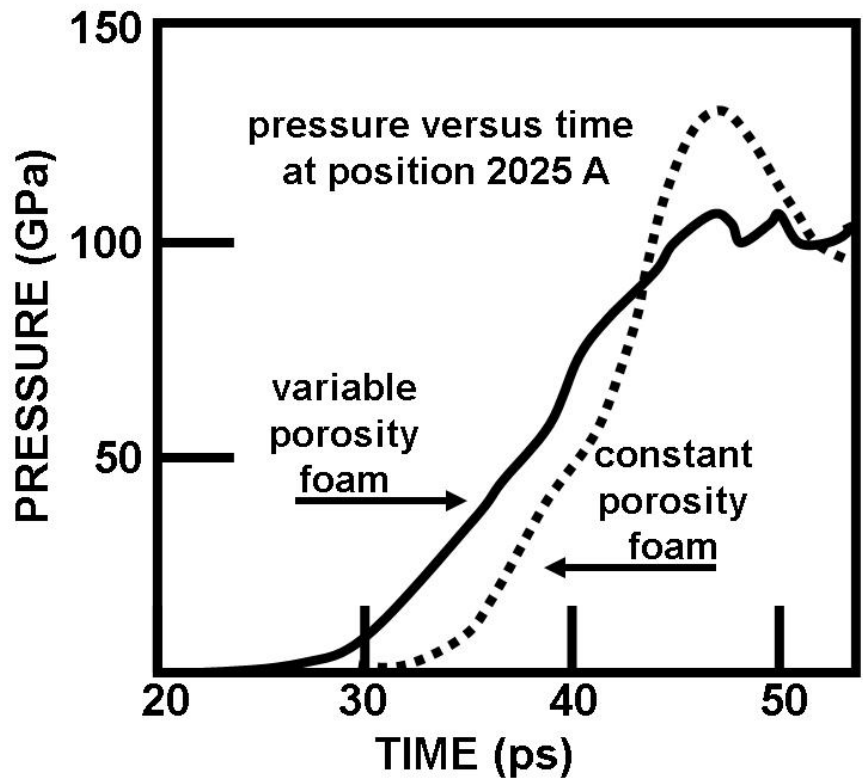

Figure 7.

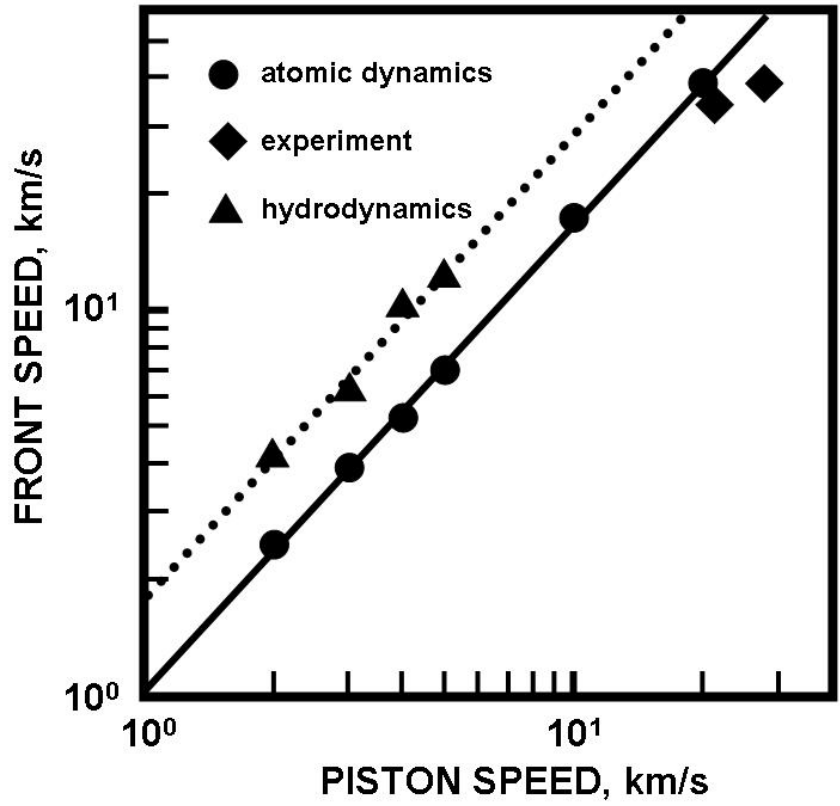